\documentclass[3p,twocolumn]{elsarticle}

\usepackage{lineno,hyperref}
\modulolinenumbers[1]

\usepackage{amsmath} %$...\textrm{normal text}...$ makes text too big in a superscript, so need \text
\usepackage{amssymb} %\lesssim

\journal{Ultramicroscopy}

\begin{document}

\begin{frontmatter}

%\title{Electron beam-induced current imaging with sub $3$-angstrom resolution}
%\title{Differential electron yield STXM contrast}
%\title{STXM contrast generated via differential electron yield}
%\title{Differential electron yield contrast generated with STXM}
\title{Differential electron yield imaging with STXM}
%\tnotetext[mytitlenote]{Fully documented templates are available in the elsarticle package on %\href{http://www.ctan.org/tex-archive/macros/latex/contrib/elsarticle}{CTAN}.}

%\author{William A. Hubbard}
%\address{Department of Physics and Astronomy, University of California, Los Angeles, CA 90095, U.S.A.}
%\address{California NanoSystems Institute, University of California, Los Angeles, CA 90095, U.S.A.}
%
%\author{Jared J. Lodico}
%\address{Department of Physics and Astronomy, University of California, Los Angeles, CA 90095, U.S.A.}
%\address{California NanoSystems Institute, University of California, Los Angeles, CA 90095, U.S.A.}
%
%\author{Xin Yi Ling}
%\address{Department of Physics and Astronomy, University of California, Los Angeles, CA 90095, U.S.A.}
%\address{California NanoSystems Institute, University of California, Los Angeles, CA 90095, U.S.A.}
%
%\author{Brian Zutter}
%\address{Department of Physics and Astronomy, University of California, Los Angeles, CA 90095, U.S.A.}
%\address{California NanoSystems Institute, University of California, Los Angeles, CA 90095, U.S.A.}
%
%\author{David Shapiro}
%\address{Advanced Light Source, Lawrence Berkeley National Laboratory, Berkeley, CA, 94720, U.S.A.}
%
%\author{B. C. Regan}
%\address{Department of Physics and Astronomy, University of California, Los Angeles, CA 90095, U.S.A.}
%\address{California NanoSystems Institute, University of California, Los Angeles, CA 90095, U.S.A.}
%\email{regan@physics.ucla.edu}

\author{William A. Hubbard$^{1,2}$, Jared J. Lodico$^{1,2}$, Xin Yi Ling$^{1,2}$, Brian Zutter$^{1,2}$, Young-Sang Yu$^{3}$, David Shapiro$^{3}$, B. C. Regan$^{1,2}$}
\address{$^{1}$Department of Physics and Astronomy, University of California, Los Angeles, CA 90095, U.S.A.}
\address{$^{2}$California NanoSystems Institute, University of California, Los Angeles, CA 90095, U.S.A.}
\address{$^3$Advanced Light Source, Lawrence Berkeley National Laboratory, Berkeley, CA, 94720, U.S.A.}
%\email{regan@physics.ucla.edu}

\begin{abstract}

Total electron yield (TEY) imaging is an established scanning transmission X-ray microscopy (STXM) technique that gives varying contrast based on a sample's geometry, elemental composition, and electrical conductivity.  However, the TEY-STXM signal is determined solely by the electrons that the beam ejects from the sample. A related technique, X-ray beam-induced current (XBIC) imaging, is sensitive to electrons and holes independently, but requires electric fields in the sample. Here we report that multi-electrode devices can be wired to produce differential electron yield (DEY) contrast, which is also independently sensitive to electrons and holes, but does not require an electric field. Depending on whether the region illuminated by the focused STXM beam is better connected to one electrode or another, the DEY-STXM contrast changes sign. DEY-STXM images thus provide a vivid map of a device's connectivity landscape, which can be key to understanding device function and failure.  To demonstrate an application in the area of failure analysis, we image a 100~nm, lithographically-defined aluminum nanowire that has failed after being stressed with a large current density.%

\end{abstract}

\begin{keyword}
\texttt{STXM, TEY, XBIC, scanning transmission X-ray microscopy, electron yield, failure analysis}
%\MSC[2010] 00-01\sep  99-00
\end{keyword}

\end{frontmatter}

%\linenumbers

\section{Introduction}

In scanning transmission X-ray microscopy (STXM), a focused X-ray beam is rastered across a thin sample, and the measured transmission is associated with the beam position to form an image. With soft (100 - 2,200 eV) X-rays, STXM offers distinct advantages over other spectromicroscopy techniques. Its sub-50~nm\cite{behyan_surface_2011,vila-comamala_advanced_2009,obst_3d_2014} spatial resolution is better than the $\sim 1\,\mu$m resolution of Raman imaging, and its beam-induced radiation damage is less that that of electron energy loss spectroscopy (EELS) in a transmission electron microscope (TEM) \cite{behyan_surface_2011}.  STXM has found broad application in the biological and physical sciences \cite{obst_3d_2014, kirz_soft_1995,warwick_scanning_1998,vyvenko_X-ray_2002,watts_soft_2009,chayanun_combining_2019} and has been used to study device physics in solar cells \cite{vyvenko_X-ray_2002,watts_soft_2009}, spin-torque memory\cite{bernstein_nonuniform_2011}, resistive memory\cite{koehl_evidence_2013} , and the Li-ion battery cathode material Li$_x$FePO$_4$\cite{lim_origin_2016}.

STXM characterizes physical structure: it determines a sample's morphology and can even spectroscopically quantify a sample's chemical composition.  However, in some cases the information returned is still too crude to identify gross characteristics of the sample that are of paramount importance.  For instance, in an electronic device two conductors might be separated by a few nanometers of insulator. Conventional STXM might identify copper on one side and aluminum on the other, but, with its limited spatial resolution, conventional STXM is ill-suited to determine whether the two conductors are electrically connected.  Because of the intimate relation between connectivity and function in electronic devices, determining the presence (or absence) and properties of such a connection might be the primary motivation for imaging the sample in the first place.

\begin{figure*}
	\begin{center} 
		{\includegraphics[width=0.9\textwidth]{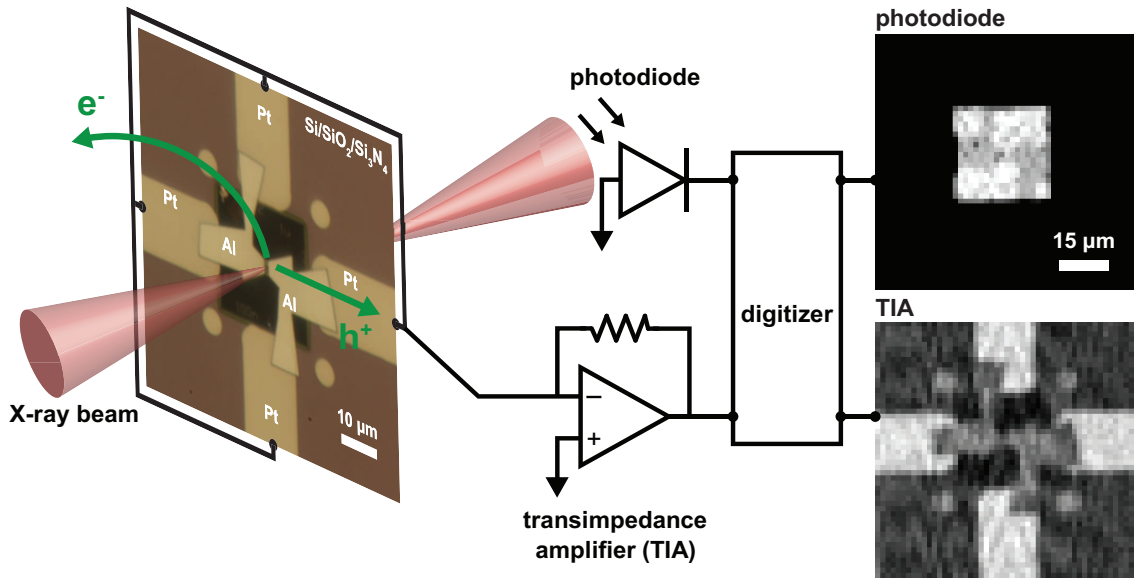}}  
		\caption{\label{diagram} \textbf{Experiment overview.} The sample (optical image on left) consists of a $200\,\mu$m-thick silicon chip supporting a 20~nm-thick silicon nitride membrane. Platinum leads over the silicon contact an aluminum pattern that tapers to an unresolved wire in the membrane's center. Here all of the Pt leads are shorted together to produce a TEY image. As the X-ray beam (red) scans the sample, the signal from the photodiode and the transimpedance amplifier (i.e. TIA, or current meter) are digitized simultaneously to form the images on the right. The photodiode signal generates the standard STXM image (top right). The TIA measures the current produced in the sample by the X-ray beam (bottom right). When the beam ejects electrons from the sample, the resulting hole current is positive and is displayed with bright contrast. } 
	\end{center}
\end{figure*}

A conventional STXM system detects the transmitted X-rays with, for example, a photodiode on the beam-exit side of the sample.  To expand its capabilities, STXM imaging techniques that instead rely on electron detection have been developed. Among the most prominent are total electron yield (TEY) and X-ray beam-induced current (XBIC) imaging. TEY is performed either by capturing electrons emitted from the sample in a remote electron detector \cite{erbil_total-electron-yield_1988,behyan_surface_2011}, or by measuring the resulting holes with a current meter attached to the sample\cite{2005Kim,behyan_surface_2011}. TEY measures beam-ejected electrons of all energies, including primary\footnote{In the X-ray microscopy community a primary electron is one scattered in a collision with beam X-ray, while in the electron microscopy community a primary electron is a beam electron, and a secondary electron is one scattered by a primary. In this article we use the conventions of the X-ray community.}, secondary, and Auger electrons\cite{frazer_probing_2003}. XBIC, on the other hand, requires a current meter attached to the sample. It measures the current generated when the X-ray beam produces electron-hole pairs that are subsequently separated by local electric fields inside the sample\cite{vyvenko_X-ray_2002,watts_soft_2009,stuckelberger_engineering_2017}.  Generally XBIC signals, where present, are larger than TEY signals, because more electron-hole pairs than ejected electrons are produced per primary X-ray.  %Despite the different processes underlying each imaging mode, %a subset of 

XBIC has an electron microscopy counterpart, (standard) electron beam-induced current (EBIC) imaging, where the electron-hole pairs are instead produced by a scanned electron beam\cite{everhart_novel_1964,leamy_charge_1982}. A related electron microscopy technique, secondary electron emission EBIC (SEEBIC) imaging\cite{hubbard_stem_2018,mecklenburg_electron_2019,hubbard_scanning_2019}, is closely analogous to TEY, and to the subject of this paper.

If the sample is wired for current collection, both TEY and XBIC imaging can be performed using the same apparatus, but with slightly different electrical connections. TEY requires only a single connection between the sample and the current meter (generally a transimpedance amplifier, or TIA)\cite{stuckelberger_engineering_2017}, while XBIC requires that the sample have an additional connection to a low impedance to allow for charge neutralization. %ground to prevent carrier pairs from recombining in the sample (producing no net current in the TIA).

Using a sample wired with multiple electrical connections, as is characteristic of XBIC and not TEY, we perform STXM mapping of electron yield. However, the resulting contrast  has its root in the ejection of electrons from the sample (and not in the creation of electron-hole pairs), as is characteristic of TEY and not XBIC. Here we report that using multiple electrodes allows differential electron yield (DEY) imaging, which gives contrast that changes sign between neighboring electrodes on the sample.  For instance, when the X-ray beam is incident on an electrode connected to the current meter, the measured current is generally positive, since the ejected electrons leave a hole current behind.  But when the beam moves to a neighboring, grounded electrode, the beam-induced hole current is shunted to ground and is therefore not measured. Meanwhile, some of the primary and secondary electrons ultimately return to the first electrode, where they are measured as a negative current (analogous to Fig.~2 of reference~\cite{hubbard_stem_2018}). This negative current represents electrons that, in the absence of the current meter, would \emph{not} have left the sample, thus by definition it is distinct from the TEY current.  The resulting DEY contrast, unlike standard STXM, TEY, or XBIC contrast, can vividly reveal whether neighboring electrodes are connected.

\begin{figure}%[!ht]
	\begin{center} 
		{\includegraphics[width=0.45\textwidth]{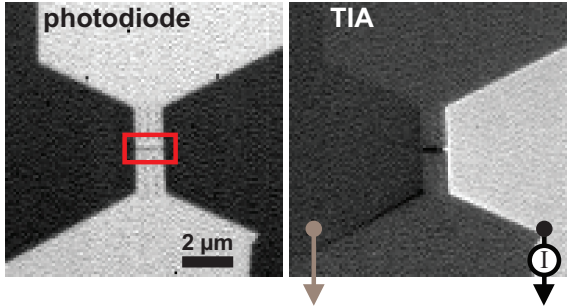}}  
		\caption{\label{oneElectrode} \textbf{STXM and DEY imaging of the Al nanowire device.} These images of the device of Fig.~\ref{diagram}  are acquired with the left electrode grounded and the right electrode attached to the TIA (indicated schematically here with an ``$I$'' circumscribed by a circle). The standard STXM image (left) shows both Al leads with the same contrast, while the DEY image (right) indicates that only the Al lead on the right is electrically connected to the TIA. The red box indicates the region shown in Fig.~\ref{ptyHiMag}.}%\textbf{Differential electron yield contrast mapping of connectivity}
	\end{center}
\end{figure}

Our implementation of DEY imaging employs a TEM sample holder, which has some particular advantages for \emph{in situ} STXM imaging of electronic devices. The production of STXM-compatible, electrically connected samples shares many challenges with the production of samples for \emph{in situ} TEM experiments. Accordingly, several X-ray beamlines have incorporated TEM stage/load-lock mechanisms in X-ray imaging systems, allowing for STXM  experiments to be performed with TEM sample holders\cite{maser_development_1998,lim_origin_2016,shapiro_cosmic_2018}.   We adopt this approach \cite{shapiro_cosmic_2018}, which gives access to the numerous off-the-shelf \emph{in situ} capabilities afforded by specialized TEM holders, including imaging in liquid and gas, heating, cooling, biasing, and physical manipulation.  The TEM stage and load-lock combination also makes for faster sample exchange (minutes instead of hours) and easier correlative TEM imaging (which can be performed without even removing the sample from the TEM sample holder).

\section{Experimental}

X-ray imaging is performed at Lawrence Berkeley National Lab’s Advanced Light Source (ALS) on beamline 7.0.1.2 (COSMIC) \cite{shapiro_cosmic_2018}.  The COSMIC beamline offers a 250--2500~eV X-ray energy range and a 50~nm spot size, and is equipped with a FEI CompuStage load-lock system, which accepts TEM sample holders. Except where indicated otherwise, STXM images are acquired with an incident beam energy of 1565~eV.  To form STXM and electron yield images, the signals from a post-sample photodiode and a FEMTO DLPCA-200 TIA, respectively, are digitized simultaneously as the beam is rastered pixel-by-pixel across the sample. To acquire diffraction patterns for ptychography, the photodiode can be retracted to expose a CCD. Scanning TEM (STEM) imaging is performed in an FEI Titan 80--300 STEM at 80~kV. For both STXM and STEM the sample is mechanically supported and electrically contacted with a Hummingbird Scientific biasing TEM sample holder. 

Our demonstration sample (Fig.~\ref{diagram} optical image) is a silicon chip patterned via optical lithography with four Ti/Pt (5/25 nm) electrodes that lead to a 20~nm-thick silicon nitride membrane\cite{hubbard_stem_2018}. On the membrane a 1-$\mu$m-long, 100-nm-wide, and 100-nm-thick Al wire is patterned via electron beam lithography. Tapered pads connect the wire to the Ti/Pt electrodes in a 4-wire configuration. Before being loaded in the STXM chamber, the wire is biased in vacuum until failure and then stored in the ambient atmosphere for several days.  AC line noise has been removed from the electron yield images via Fourier filtering, and current values are given relative to the signal on the bare silicon nitride membrane, where very little XBIC is expected. The optical density referenced in Figs.~\ref{energyStack} and \ref{energyPlots} is $-\ln \frac{I}{I_0}$, where $I_0$ is the photodiode signal on the bare silicon nitride membrane, and is filtered by principal component analysis \cite{lerotic_cluster_2004}.% $-log_{10}\frac{I}{I_o}$ before extracting the spectra

\section{Results and Discussion}

STXM imaging of the silicon nitride membrane window reveals the Al electrodes, which transmit fewer photons than the bare membrane and thus appear slightly darker (Fig.~\ref{diagram} top right). But STXM imaging of the silicon support frame provides no information, as the opaque sample blocks the incident X-rays. The (total) electron yield image, on the other hand, reveals device features in the entire field of view, even where the sample is opaque (Fig.~\ref{diagram} bottom right). The Al pads are visible, as in the STXM image, but so are the Pt electrodes to which the Al is connected. The Pt has a larger electron yield than the Al and therefore appears brighter. Four Pt islands at the corners of the membrane are also visible, despite the apparent lack of an electrical connection.  Holes produced in these islands can evidently travel the several-micrometer distance to the Pt electrodes \cite{hubbard_stem_2018}. Contrast is slightly darker over the membrane, an insulator that generates few primary electrons in the beam. 

\begin{figure}[!t]
	\begin{center} 
		{\includegraphics[width=0.45\textwidth]{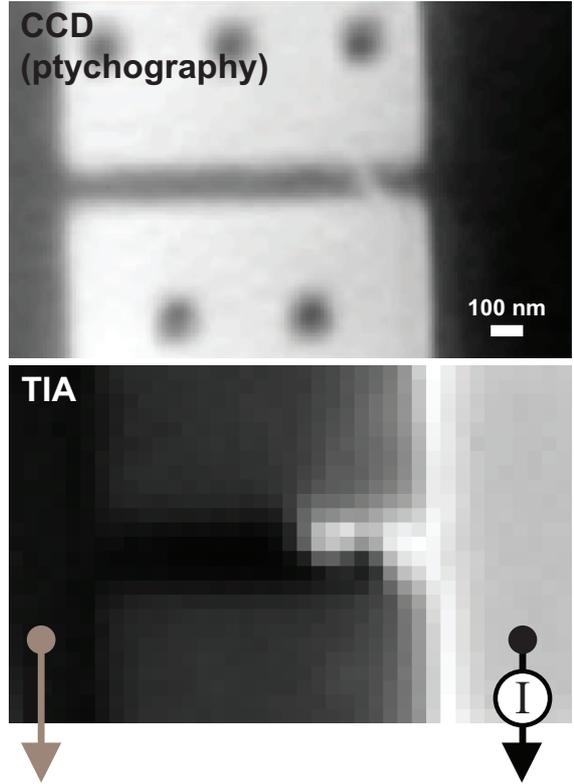}}  
		\caption{\label{ptyHiMag} \textbf{Ptychography and DEY imaging of the Al nanowire device.} Retracting the photodiode and scanning over the region outlined in red in Fig.~\ref{oneElectrode} produces, after reconstruction, a ptychography image (top) that reveals the break in the Al nanowire. The simultaneously acquired electron yield image (bottom) has the inferior resolution, relative to ptychography, of standard STXM, but it nonetheless reveals a surprising feature: electrical connectivity spans the `break' in the Al wire that is seen in ptychographic image.}
	\end{center}
\end{figure}

\begin{figure*}%[!ht]
	\begin{center} 
		{\includegraphics[width=\textwidth]{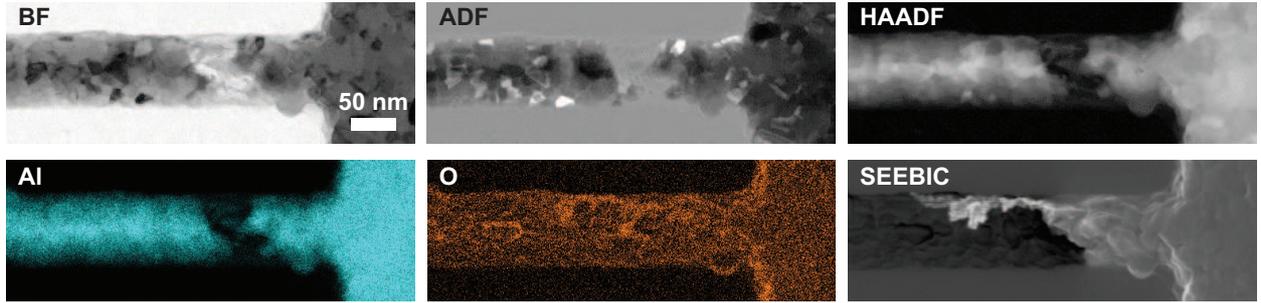}}  
		\caption{\label{STEMforXBIC} \textbf{STEM imaging of the Al nanowire device.} The Al wire of Figs.~\ref{diagram}--\ref{ptyHiMag} is imaged with standard STEM (BF, ADF, and HAADF), STEM EDS elemental mapping (Al and O), and STEM SEEBIC. The BF and SEEBIC images are the electron microscopy analogues of the previously-shown STXM (Fig.~\ref{oneElectrode}) and DEY images (Figs.~\ref{oneElectrode}--\ref{ptyHiMag}) respectively. The STEM images show similar contrast but significantly better spatial resolution relative to their analogous X-ray images.}
	\end{center}
\end{figure*}

Electron yield mapping can be extremely helpful in samples that are mostly opaque. With only the transmission-based contrast of standard STXM, locating a thin region is generally accomplished by trial-and-error, and is analogous to wandering around in the dark. Electron yield imaging turns the lights on: sample features far from the transparent area can be used as landmarks to locate the region of interest systematically and quickly.

The device of Fig.~\ref{diagram} features an unresolved Al wire that previously connected the two larger pads. Because the device has been subjected to a bias current sufficiently large to cause heating and eventual failure, the wire is broken and represents a very large electrical impedance.  We image the nanowire of Fig.~\ref{diagram} again, this time at higher magnification (Fig.~\ref{oneElectrode}), but here we change the electrical connections for DEY imaging:  the right Al electrode remains connected to the TIA but the left electrode is now grounded. (The biasing sample holder gives independent access to each of the four Ti/Pt electrodes, so this change can be made without breaking vacuum.) 

In this configuration, when the X-ray beam ejects electrons from the right electrode, the TIA measures a positive (hole) current. When the X-ray beam ejects electrons from the left electrode, the hole current flows to ground directly and is not measured by the TIA. However, a fraction of the electrons emitted from the left electrode are recaptured\cite{hubbard_stem_2018} by the right electrode and are measured as a negative (electron) current. Thus, the resulting image (Fig.~\ref{oneElectrode} right) shows each electrode as bright or dark respectively, depending on whether or not the electrode is directly connected to the TIA. Like TEY, DEY imaging maps whether or not a region is conducting:  the Al on both sides of the break more readily emits primary electrons than the insulating Si$_3$N$_4$ support membrane.  But DEY imaging also indicates the connectivity landscape, particularly the `watershed' boundary of the region electrically connected to the TIA \cite{hubbard_stem_2018}. Such differential contrast is not accessible with TEY.

Note that the dark contrast generated by electron recapture (e.g. the left electrode of Fig.~\ref{oneElectrode} right) indicates that DEY imaging,  on electrodes showing bright contrast (e.g. the right electrode of Fig.~\ref{oneElectrode} right), always has a better signal-to-noise ratio than TEY imaging.  The recaptured electron current has the opposite sign as the hole current.  To the extent that these currents are equal and are collected by the same TIA, they cancel.  Viewed from this perspective, TEY is a worst case scenario, in that the recapturing electrode spans the whole sample. It thus collects a correspondingly large recapture current, and generates a correspondingly small net current (i.e. signal). One can even imagine pathological geometries where a nearby, off-sample surface, such as an aperture \cite{2005Kim}, could produce enough primary and secondary electrons --- which contain no information about the sample itself --- to overwhelm the original hole current. Imaging a small electrode that alone is connected to the TIA gives the best case scenario, for here the recapture current is minimized and the measured hole current is undiminished.

Still higher magnification scans of the same device (Fig.~\ref{ptyHiMag}) resolve both the physical and the electronic break in the Al wire. Here we retract the photodiode to capture the diffraction pattern generated at each X-ray beam position (i.e. pixel) for ptychography. Without the photodiode the standard STXM image is no longer available. Ptychographically reconstructing the captured diffraction patterns produces an image that reveals a break in the Al on the right side of the wire (Fig.~\ref{ptyHiMag} top). The break appears clean, with an $\sim 50$~nm length missing from the wire.   The DEY image (Fig.~\ref{ptyHiMag} bottom), however, shows a more complicated structure around the break.  The large Al lead on the right is bright, as expected based on the lower magnification image of the same device (Fig.~\ref{oneElectrode} right). But surprisingly, portions of the wire to the left of the `break' (as identified by the ptychographic image) are also bright, indicating that they too are connected to the Al lead on the right.

During ptychographic imaging, the photodiode is retracted and thus its signal is not available.  However, electron yield data can still be acquired simultaneously with the diffraction patterns used to produce the ptychographic image.  And unlike the ptychographic data, the electron yield data is immediately viewable in a real-space format without any reconstruction.  The real-time feedback provided by electron yield imaging, like the ability to image opaque regions of a sample, is an experimental convenience that can save valuable time on the beamline.

\begin{figure}%[!ht]
	\begin{center} 
		{\includegraphics[width=0.45\textwidth]{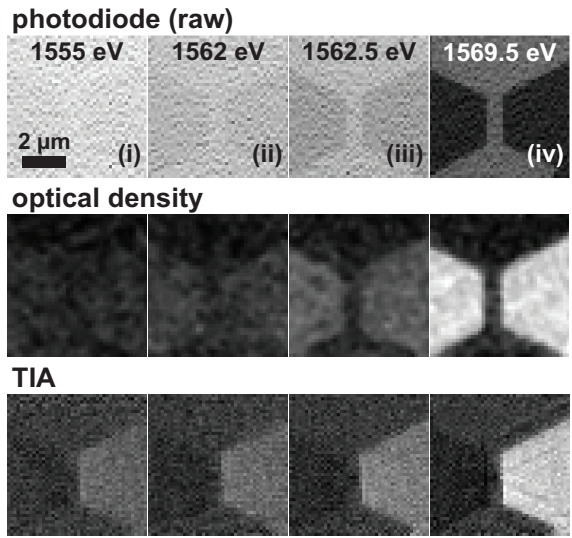}}  
		\caption{\label{energyStack} \textbf{STXM and electron yield images at four representative X-ray beam energies.} The beam energy for each column of representative images (see Fig.~\ref{energyPlots}) is indicated.   The electrodes are almost invisible in the raw photodiode (upper row) and calculated optical density (middle row) images below 1562~eV, while they are easily seen in the electron yield images (bottom row) over the entire energy range scanned (1555--1575~eV). The electron yield images are acquired with the circuit as indicated in Figs.~\ref{oneElectrode}--\ref{ptyHiMag}. The contrast scale is held fixed for each row of images. }
	\end{center}
\end{figure}

\begin{figure}%[ht!]
	\begin{center} 
		{\includegraphics[width=0.45\textwidth]{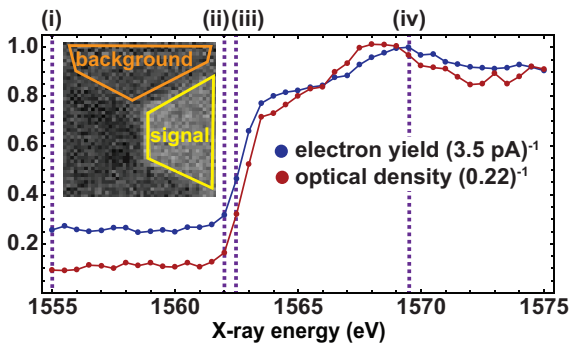}}  
		\caption{\label{energyPlots} \textbf{Electron yield and optical density of an Al electrode as a function of incident beam energy.} Signal on the right electrode (inset, yellow) is plotted relative to the background reference region (inset, orange)  for the electron yield (blue curve) and optical density (red curve). Dashed lines indicate images shown in Fig.~\ref{energyStack}.}
	\end{center}
\end{figure}

The use of the TEM sample holder for X-ray imaging makes correlative microscopy especially straightforward. STEM (Fig.~\ref{STEMforXBIC}) imaging of the same device in the same sample holder confirms, with much improved spatial resolution, the device properties ascertained with X-ray imaging. Bright-field (BF), annular dark-field (ADF), and high-angle ADF (HAADF) STEM images each show loss of material at the failure point, and energy-dispersive X-ray spectroscopy (EDS) elemental mapping confirms that Al has disappeared in the gap. SEEBIC imaging \cite{hubbard_stem_2018} shows the same non-obvious electrical connectivity seen with DEY imaging, again with improved spatial resolution: the right electrode is electrically connected to material well to the left of the gap that appears in the standard imaging channels.  Metallic aluminum in quantities below the detection limits here is likely responsible for this connectivity extension. Some correlation between the connectivity extension seen with DEY and SEEBIC imaging is seen in the oxygen EDS map, but nothing that would suggest the existence of the extension without the DEY (or SEEBIC) data.  In many practical situations, DEY's ability to detect the electrical connectivity created by dopants or other trace impurities in quantities below the standard detection methods' thresholds might be key to understanding device behavior.

In X-ray microscopy, unlike TEM, the beam energy can be tuned across an absorption threshold of an element in the sample. The differential contrast in the electron yield persists under such spectroscopic imaging. We scan the beam energy over 41 values encompassing the aluminum K-edge (1555~eV to 1575~eV in 0.5~eV steps).  Below 1562~eV, the Al electrodes are difficult to detect in the STXM images, while they are obvious in the electron yield images (Fig.~\ref{energyStack}).  Both signals become more intense (Fig.~\ref{energyPlots}) as the energy exceeds the Al K-edge threshold at $\sim$1563~eV. The Al electron yield, which is already significant below the K-edge, increases by about 400\% immediately above the K-edge.%,

\section{Conclusion}

We have demonstrated STXM electron yield imaging of a simple device mounted in a TEM biasing holder. With a TEM load-lock installed, performing electron yield measurements requires no modification of the STXM chamber or the data acquisition electronics; all electrical connections to the device are made through the holder, and the electron yield signal is digitized in parallel with the existing photodiode signal.  Measuring current from the entire device provides the standard TEY measurement, while grounding portions of the circuit gives DEY images that map connectivity within the device. In a broken Al nanowire, the differential contrast provided by DEY imaging precisely locates the failure point and reveals a non-obvious electrical connection spanning the physical gap in the wire.  As a complement to standard STXM and ptychographic imaging, the DEY technique has a number of practical advantages, including real-time and opaque-region imaging. For functional studies of micro- and nano-scale electronic devices, DEY imaging makes a particularly powerful addition to the suite of available correlative imaging modes.%

\section{Acknowledgments}
This work was supported by National Science Foundation (NSF) Science and Technology Center (STC) award DMR-1548924 (STROBE), by NSF award DMR-1611036, and by the UCLA PSEIF.  Work at the ALS was supported by the Office of Science, Office of Basic Energy Sciences, of the US Department of Energy under contract number DE-AC02-05CH11231. The authors acknowledge the use of instruments at the Electron Imaging Center for NanoMachines supported by NIH 1S10RR23057 and the CNSI at UCLA.  The authors also thank M. Murnane and J. Miao for encouragement and assistance in arranging the experiments.

\bibliography{xbicBib_v3_delete_longUrl}

\end{document}